\documentclass[prd,aps,showpacs,preprintnumbers,amssymb]{revtex4}

\usepackage{color}
\usepackage{epsf}

\def\e3p{$\eta \rightarrow 3 \pi$}

\begin{document}
\title{%
\hfill{\normalsize\vbox{%
\hbox{}
 }}\\
{Standard model effective potential from trace anomalies}}

\author{Renata Jora
$^{\it \bf a}$~\footnote[1]{Email:
 rjora@theory.nipne.ro}}

\affiliation{$^{\bf \it a}$ National Institute of Physics and Nuclear Engineering PO Box MG-6, Bucharest-Magurele, Romania}

\date{\today}

\begin{abstract}
By analogy with the low energy QCD effective linear sigma model we construct a standard model effective potential based entirely on the requirement that the tree level and quantum level trace anomalies must be satisfied. We discuss a particular realization of this potential in connection to the Higgs boson mass and Higgs boson effective couplings to two photons and two gluons. We find that this kind of potential may describe well the known phenomenology of the Higgs boson.

\end{abstract}
\maketitle

\section{Introduction}

With the discovery of the electroweak Higgs boson by the Atlas \cite{Atlas} and CMS \cite{CMS} experiments the standard model has entered an era of unprecedented experimental confirmation with few hints with regard to its possible extensions to accommodate other particles, interactions or symmetries. Even in its early years the standard model Higgs boson has been the subject of a flurry of theoretical papers that dealt with its properties
\cite{Glashow}-\cite{Hagen}, the effective one or two loops potential \cite{Sher}, \cite{Jones1} \cite{Jones2}, naturalness of the electroweak scale \cite{Veltman} or the vacuum stability of the standard model \cite{Casas1}, \cite{Casas2}, \cite{Espinosa}.

It is relatively straightforward to compute  the effective potential for a theory with spontaneous symmetry breaking \cite{Jones1},\cite{Jones2}.  This potential then may be renormalization group improved \cite{Casas1}, \cite{Casas2}, constrained to be scale invariant and the associated vacuum expectation value or effective mass computed.

Historically the electroweak model with spontaneous symmetry breaking and the $SU(2)_L \times SU(2)_R$ linear sigma model for low energy QCD have been strongly related.  The latter also displays spontaneous symmetry breaking associated to the formation of quark condensates and possessed also three Goldstone bosons, the pions. However the QCD linear sigma model was not straightforwardly derived in some loop  order from the more basic QCD as the hadron detailed structure is of as yet unknown but rather based on the specific properties and symmetries already observed in the hadron spectrum. It is worth mentinoning that linear sigma models have long been a basic tool for some effective description for low energy QCD  and were generalized to the more comprehensive global group $SU(3)_L \times SU(3)_R$ \cite{Schechter1}-\cite{Schechter4}  and also to include four quark states \cite{Jora1}-\cite{Jora5} that can accommodate two scalar and two pseudoscalar nonets. Moreover without the specific knowledge of the detailed interaction one can add phenomenological terms that mock up the axial \cite{Jora6} and the trace anomalies \cite{Jora7} with significant role in the hadron properties and good agreement with the experimental data.

In this work we will construct an effective potential based entirely on the trace anomaly terms at tree and quantum level.  First in section II we will propose a general version where the parameters are constrained only by the requirement of mocking up exactly the trace anomaly. Then the model will contained a number of unknown parameters which should be determined from the phenomenological data.  Further on in section III based on the analogy with low energy QCD   we introduce a particular version of the same potential where all the parameters are specified. We will study  in this context the minimum equations and  the mass of the Higgs boson. In section IV we analyze in the same framework the Higgs effective couplings to two photons and two gluons, relevant for the associated decays. Section V is dedicated to Conclusions.

\section{Trace anomaly induced potential}
We start by considering the relevant part of the standard model Lagrangian apart for the kinetic terms for the fermions and for the Higgs doublet. Our choice is motivated by the fact that these terms are scale invariant at the tree level and for the quantum renormalized Lagrangian there is no contribution to the trace anomaly since there is no coupling constant in front of these terms. The gauge fields however behave differently; we can always transform the gauge field as $gA^a_{\mu}\rightarrow A^a_{\mu}$ where $A^a_{\mu}$ is generic arbitrary gauge field and $g$ is its coupling constant. Then the corresponding kinetic term in the Lagrangian will appear with a factor $\frac{1}{g^2}$ which will contribute to the trace anomaly through its beta function.  The relevant part of the Lagrangian is then:
\begin{eqnarray}
{\cal L}_s=-\frac{1}{4g^2}F^{a\mu\nu}F^a_{\mu\nu}-\frac{1}{4g^{\prime2}}G^{\mu\nu}G_{\mu\nu}+(y\bar{q}_L\tilde{{\bf \Phi}}t_R+h.c.)-\frac{m^2}{2}{\bf \Phi}^{\dagger}{\bf \Phi}-\frac{\lambda}{6}({\bf \Phi}^{\dagger}{\bf \Phi})^2,
\label{relalgr00}
\end{eqnarray}
where ${\bf \Phi}$ is the Higgs doublet and,
\begin{eqnarray}
&&F_{\mu\nu}^a=\partial_{\mu}A^a_{\nu}-\partial_{\nu}A^a_{\mu}+g\epsilon^{abc}A^b_{\mu}A^c_{\nu}
\nonumber\\
&&G_{\mu\nu}=\partial_{\mu}B_{\nu}-\partial_{\nu}B_{\mu}
\label{notfield7564}
\end{eqnarray}
and $F^a_{\mu\nu}$ is the $SU(2)$ field tensor and $G_{\mu\nu}$ is the $U(1)_Y$ one. Moreover since  all fermions except for the top quark have small masses compared to the electroweak scale we considered only the term pertaining to the top quark and its associated left handed doublet. By definition all the terms in Eq. (\ref{relalgr00}) are gauge invariant.

 The next step is  to take into account all trace anomalies known at both tree and  quantum levels. It is known that the mass terms break scale invariance at tree level. However the quantum breaking of the scale transformation deserves a more detailed discussion.  The trace anomaly refers to the renormalized Lagrangian. Then for a general Lagrangian depending on the fields $\Phi_i$ (fermions or bosons) and coupling constants $\lambda_i$ the quantum corrections to the trace anomalies are given by:
\begin{eqnarray}
\theta^{\mu}_{\mu}=\Bigg[\sum_i\frac{\partial {\cal L}}{\partial \Phi_i}\frac{\partial \Phi_i}{\partial \sigma}+\sum_i\frac{\partial{\cal L}}{\partial \lambda_i}\frac{\partial \lambda_i}{\partial \sigma}\Bigg]\sigma,
\label{trscae43443}
\end{eqnarray}
where $x'=\exp[\sigma]x$ and $\sigma$ is the scale associated to the scale transformation. One may write,
\begin{eqnarray}
\sum_i\frac{\partial {\cal L}}{\partial \lambda_i}\frac{\partial \lambda_i}{\partial \sigma}\sigma=
\frac{\partial {\cal L}}{\partial \lambda_i}\beta(\lambda_i).
\label{secterm466564}
\end{eqnarray}
Next we observe that in functional sense as it is the case:
\begin{eqnarray}
\frac{\partial{\cal L}}{\partial\Phi_i}|_f=\frac{\partial {\cal L}}{\partial \Phi_i}-\partial_{\mu}(\frac{\partial {\cal L}}{\partial_{\mu}\Phi_i}),
\label{motrr645665}
\end{eqnarray}
which is zero by the equation of motion (the subscript $f$ indicates in the functional sense). Since the trace anomaly calculations implicitly assume that the equation of motion for the renormalized field is satisfied  then clearly $\frac{\partial {\cal L}}{\partial\Phi_i}|_f=0$.

Consequently only the terms that contain coupling constants contribute to the trace anomaly whereas the contribution from the dependence of the renormalized fields with the scale is cancelled by the equation of motion. For a general gauge theory with fermions and scalars one can make from the beginning the change of variable $A^a_{\mu}g\rightarrow A^a_{\mu}$ such that the coupling constant is eliminated from all gauge covariant derivatives. Thus the gauge invariant kinetic terms of the matter fermions or bosons bring no contribution to the trace anomaly.

 We shall start with the $U(1)_Y$ gauge group that we will analyze in detail and just write down the results for $SU(2)_L$ and $SU(3)_C$ that can be easily obtained by applying the same procedure. All our calculations and definitions are inspired by the work in \cite{Schechter1}-\cite{Schechter5} by analogy with low energy QCD. Thus for a generic Lagrangian of the type \cite{Schechter5},
\begin{eqnarray}
{\cal L}=-\frac{1}{2}\partial^{\mu}\eta\partial_{\mu}\eta -V(\eta),
\label{schlagr54664}
\end{eqnarray}
the new improved energy momentum tensor is defined as:
\begin{eqnarray}
\theta_{\mu\nu}=\delta_{\mu\nu}{\cal L}+\partial_{\mu}\eta\partial_{\nu}\eta-\frac{1}{6}(\partial_{\mu}\partial_{\nu}-\delta_{\mu\nu}\Box)\eta^2.
\label{tensr4554}
\end{eqnarray}
This leads upon applying the equation of motion for the field $\eta$ to the following trace of the energy momentum tensor:
\begin{eqnarray}
\theta_{\mu\mu}=\eta\frac{\partial V}{\partial \eta}-4V.
\label{treta54655}
\end{eqnarray}
We will apply   Eq. (\ref{treta54655}) consistently in all our subsequent calculations of course adjusted to the specific Lagrangian.

The trace anomaly for $U(1)_Y$ reads \cite{Schechter10}, \cite{Peskin}:
\begin{eqnarray}
\theta^{\mu}_{\mu}=\frac{\beta(g')}{2g^{\prime 3}}G^{\mu\nu}G_{\mu\nu}.
\label{tecv98}
\end{eqnarray}
We rescale the gauge fields back to their original form $B_{\mu}\rightarrow B_{\mu}g'$ which leads to:
\begin{eqnarray}
\theta^{\mu}_{\mu}=\frac{\beta(g')}{2g^{\prime }}G^{\mu\nu}G_{\mu\nu}.
\label{tecv9877}
\end{eqnarray}
Then we consider $\Phi$ a slowly varying background Higgs field and introduce the term:
\begin{eqnarray}
V_1=b_1G^{\mu\nu}G_{\mu\nu}\ln[\frac{x_1G^{\mu\nu}G_{\mu\nu}}{\Lambda^4}]+b_2G^{\mu\nu}G_{\mu\nu}\ln[\frac{y_1\Phi^4}{\Lambda^4}],
\label{firdtermpot645}
\end{eqnarray}
where $\Lambda$ is some arbitrary scale and $b_1$, $b_2$, $x_1$ and $x_2$ are arbitrary dimensionless coefficients. We compute the trace of the energy momentum tensor for the potential in Eq. (\ref{firdtermpot645}) as:
\begin{eqnarray}
\theta^{\mu}_{\mu}=\frac{\partial V_1}{\partial (G^{\mu\nu}G_{\mu\nu})}4G^{\mu\nu}G_{\mu\nu}+\frac{\partial V_1}{\partial \Phi}\Phi-4V,
\label{comp7675}
\end{eqnarray}
to determine that it is,
\begin{eqnarray}
\theta^{\mu}_{\mu}=(4b_1+4b_2)G^{\mu\nu}G_{\mu\nu},
\label{restr546}
\end{eqnarray}
which leads to the contraint $4b_1+4b_2=\frac{\beta(g')}{g'}$. Then we apply the equation of motion,
\begin{eqnarray}
\frac{\partial V_1}{\partial (G^{\mu\nu})}=2\Bigg[b_1G_{\mu\nu}\ln[\frac{x_1G^{\mu\nu}G_{\mu\nu}}{\Lambda^4}]+b_1G_{\mu\nu}+b_2G_{\mu\nu}\ln[\frac{y_1\Phi^4}{\Lambda^4}]\Bigg]=0
\label{mot7529}
\end{eqnarray}
 and extract the field $G^{\mu\nu}$ from the potential $V_1$:
 \begin{eqnarray}
 G^{\mu\nu}G_{\mu\nu}=\frac{y_1}{x_1}\Phi^4\exp\Bigg[-1-(\frac{b_2}{b_1})\ln[\frac{y_1\Phi^4}{\Lambda^4}]\Bigg].
 \label{extr657489}
 \end{eqnarray}
 We introduce the result in Eq. (\ref{extr657489}) into the expression for the potential in Eq. (\ref{firdtermpot645}) to obtain:
 \begin{eqnarray}
 V_1=-b_1\frac{y_1}{x_1}\Phi^4\exp\Bigg[-1-(\frac{b_2}{b_1}+1)\ln[\frac{y_1\Phi^4}{\Lambda^4}]\Bigg].
 \label{res635253}
 \end{eqnarray}
A similar expression can be determined for $SU(2)_L$:
\begin{eqnarray}
V_2=-c_1\frac{y_2}{x_2}\Phi^4\exp\Bigg[-1-(\frac{c_2}{c_1}+1)\ln[\frac{y_2\Phi^4}{\Lambda^4}]\Bigg],
\label{v2res65}
\end{eqnarray}
where $4c_1+4c_2=\frac{\beta(g)}{2g}$, $y_2$ and $x_2$ are arbitrary dimensionless coefficients  and $g$ is the weak coupling constant.
Then the potential induce by $SU(3)_C$ is just:
\begin{eqnarray}
V_3=-d_1\frac{y_3}{x_3}\Phi^4\exp\Bigg[-1-(\frac{d_2}{d_1}+1)\ln[\frac{y_3\Phi^4}{\Lambda^4}]\Bigg],
\label{v3899}
\end{eqnarray}
with $4d_1+4d_2=\frac{\beta(g_3)}{2g_3}$, $y_3$ and $x_3$  arbitrary dimensionless coefficients and  $g_3$  the strong coupling constant.

One can associate to the trace anomaly corresponding to the top Yukawa term in the Lagrangaian the potential:
\begin{eqnarray}
&&V_4=k_1(\bar{\Psi}_l\tilde{{\bf \Phi}}t_R+h.c)\ln[\frac{(x_4(\bar{\Psi}_L\tilde{{\bf\Phi}}t_R+h.c))}{\Lambda^4}]+
\nonumber\\
&&k_2(\bar{\Psi}_l\tilde{{\bf \Phi}}t_R+h.c)\ln[\frac{y_4\phi^4}{\Lambda^4}],
\label{fourtht565}
\end{eqnarray}
where the anomaly requires that $4k_1+4k_2=\frac{\beta(y)}{\sqrt{2}}$ where $y$ is the top Yukawa coupling. Here again $k_1$ and $k_2$ are arbitrary dimensionless coefficients.

The most interesting and complicated term to evaluate is however that of the mass of the Higgs bosons in conjunction with that of the quadrilinear coupling $\lambda$. The mass term is not scale invariant already at
tree level and the trace of the energy momentum tensor will receive corrections also at the quantum level. A suitable potential is then:
\begin{eqnarray}
&&V_5=\frac{1}{2}(m^2-\frac{\beta(m^2)}{2}){\bf \Phi}^{\dagger}{\bf \Phi}+
\nonumber\\
&&r_1[{\bf \Phi}^{\dagger}{\bf \Phi}]^2\ln[\frac{x_5[{\bf \Phi}^{\dagger}{\bf \Phi}]^2}{\Lambda^4}]+
r_2[{\bf \Phi}^{\dagger}{\bf \Phi}]^2\ln[\frac{y_5\Phi^4}{\Lambda^4}].
\label{res645376}
\end{eqnarray}
Here ${\bf \Phi}$ is the regular Higgs doublet and $r_1$ and $r_2$ are arbitrary dimensionless coefficients. First term gives the correct mass anomaly and the second and third terms give the correct $\lambda$ anomaly provided that $4r_1+4r_2=\frac{\beta(\lambda)}{24}$. The equation of motion $\frac{\partial V_5}{\partial {\bf\Phi}}=0$ leads to:
\begin{eqnarray}
&&2r_1[{\bf \Phi}^{\dagger}{\bf \Phi}]\ln[\frac{x_5[{\bf \Phi}^{\dagger}{\bf \Phi}]^2}{\Lambda^4}]+
\nonumber\\
&&2r_1[{\bf \Phi}^{\dagger}{\bf \Phi}]+2r_2[{\bf \Phi}^{\dagger}{\bf \Phi}]\ln[\frac{y_5\Phi^4}{\Lambda^4}]+\frac{1}{2}(m^2-\frac{\beta(m^2)}{2})=0.
\label{v5rre664}
\end{eqnarray}
Eq. (\ref{v5rre664}) is a transcendental equation. To solve it we first make the notations:
\begin{eqnarray}
&&X=[{\bf \Phi}^{\dagger}{\bf \Phi}]
\nonumber\\
&&a=4d_1
\nonumber\\
&&b=2d_1\ln[\frac{x_5}{\Lambda^4}]+2d_2\ln[\frac{y_5\Phi^4}{\Lambda^4}]+2d_1
\nonumber\\
&&c=\frac{1}{2}(m^2-\frac{\beta(m^2)}{2})
\label{not75676}
\end{eqnarray}
Then Eq. (\ref{v5rre664}) may be rewritten  as:
\begin{eqnarray}
aX\ln[X]+bX+c=0.
\label{res54664}
\end{eqnarray}
We denote $Y=\exp[\frac{b}{a}]X$ to determine:
\begin{eqnarray}
Y=W[-\frac{c}{a}\exp[\frac{b}{a}]],
\label{lambert45}
\end{eqnarray}
where $W(x)$ is the Lambert function. We can assume $x=-\frac{c}{a}\exp[\frac{b}{a}]$ small (make the final choice of the coefficients as such) case in which $W(x)\approx x$. This leads to:
\begin{eqnarray}
X=\exp[-\frac{b}{a}-\frac{c}{a}\exp[\frac{b}{a}]].
\label{finarestyy}
\end{eqnarray}
We introduce Eq. (\ref{finarestyy}) into Eq. (\ref{res645376}) to determine,
\begin{eqnarray}
V_5=c\exp[-\frac{b}{a}]-\frac{a}{4}\exp[-\frac{2b}{a}],
\label{pot56677}
\end{eqnarray}
or by using Eq. (\ref{not75676}):
\begin{eqnarray}
&&V_5=\frac{1}{2}(m^2-\frac{\beta(m^2)}{2})\frac{\sqrt{y_5}}{\sqrt{x_5}}\exp\Bigg[-\frac{1}{2}-\frac{1}{2}(1+\frac{r_2}{r_1}\ln[\frac{y_5\Phi^4}{\Lambda^4}]\Bigg]-
\nonumber\\
&&d_1\frac{y_5}{x_5}\Phi^4\exp\Bigg[-1-(1+\frac{r_2}{r_1})\ln[\frac{y_5\Phi^4}{\Lambda^4}]\Bigg].
\label{res54666383}
\end{eqnarray}
The full potential is then:
\begin{eqnarray}
V=V_1+V_2+V_3+V_4+V_5+\frac{1}{24}\lambda\Phi^4.
\label{finres545343}
\end{eqnarray}
Note that we could safely introduce the $\lambda$ term because it is scale invariant.

\section{Trace anomaly inspired particular potential}

The effective potential in Eq. (\ref{finres545343}) is constructed by analogy with low energy QCD effective models and contains in its most general form 20 parameters and 5 constraints.  In order to substantiate that nevertheless this is a good phenomenological model we need to stress out three important points: 1) The effective potential built here is an all orders potential and even if it contains a proliferation of parameters, these parameters encapsulate the intrinsic dependence on higher order loops without making any explicit calculations. Since computing beta functions and anomalous dimensions is far more amenable than calculating an effective potential or other processes at the same  loop order the apparent complexity leads in essence to an effective simplification. 2) The potential in Eq. (\ref{finres545343}) is completely independent of the nature elementary or composite of the Higgs boson and it is a reliable description also for the case when some strong dynamics is at play in the electroweak symmetry breaking. Note that the composite scenario is not completely excluded by the LHC or other experimental data \cite{PDG}. 3) The number of parameters may be greatly reduced by making an educated guess of some of the parameters by analogy with low energy QCD \cite{Schechter10} or even with the standard construction of the Higgs one loop effective potential as described in the literature \cite{Casas1}, \cite{Casas2}. Thus one can choose from physical arguments related to the relative renormalization of the wave function of the Higgs field the following expression for the constrained parameters:

\begin{eqnarray}
&&b_1=\frac{1}{4}[\frac{\beta(g')}{2g'}-\frac{\gamma}{2}]
\nonumber\\
&&c_1=\frac{1}{4}[\frac{\beta(g)}{2g}-\frac{\gamma}{2}]
\nonumber\\
&&d_1=\frac{1}{4}[\frac{\beta(g_3)}{2g_3}-\frac{\gamma}{2}]
\nonumber\\
&&b_2=c_2=d_2=\frac{\gamma}{8}
\nonumber\\
&&k_1=\frac{1}{4}[\frac{\beta(y)}{\sqrt{2}}-\frac{\gamma y}{\sqrt{2}}]
\nonumber\\
&&k_2=\frac{1}{4}\frac{\gamma y}{\sqrt{2}}
\nonumber\\
&&r_1=\frac{1}{4}[\frac{\beta(\lambda)}{24}-\frac{4\gamma\lambda}{24}]
\nonumber\\
&&r_2=\frac{1}{4}\frac{4\gamma\lambda}{24}.
\label{seccoeff567}
\end{eqnarray}
Here $\beta(g')$, $\beta(g)$, $\beta(g_3)$, $\beta(y)$ and $\beta(\lambda)$ are the beta functions for the coupling constants of the $U(1)_Y$, $SU(2)_L$, $SU(3)_c$ groups, the quadrilinear term in the Higgs potential and the top Yukawa coupling respectively (we  use the results in \cite{Jones1}). Moreover $\gamma$ is the anomalous dimension of the Higgs field.

Moreover the parameters $x_i$ and $y_i$ are redundant because they are already associated with a factor in front of the respective terms so they may be chosen as:
\begin{eqnarray}
&&x_1=y_1=\frac{1}{4}
\nonumber\\
&&x_2=y_2=\frac{1}{4}
\nonumber\\
&&x_3=y_3=\frac{1}{4}
\nonumber\\
&&x_4=y_4=\frac{y}{\sqrt{2}}
\nonumber\\
&&x_5=y_5=\frac{\lambda}{24}.
\label{first54553}
\end{eqnarray}
Here we took into account as arguments of the logarithm the natural expressions of the scalar polynomials as they appear in the Lagrangian.

Next we will set $\Phi$ constant and apply the standard approach for constructing effective potentials. We denote \cite{Casas1}, \cite{Casas2}:
\begin{eqnarray}
\Phi(t)=\xi(t)\Phi\label{den6574}
\end{eqnarray}
where t is the running parameter $\mu(t)=m_Z\exp[t]$ and,
\begin{eqnarray}
\xi(t)=\exp[-\int_0^t \gamma(t')dt'].
\label{def54829}
\end{eqnarray}
Note that for $\mu(t)=m_Z$,  $t_Z=0$.  We consider all the parameters computed at this scale and
apply the minimum equation:
\begin{eqnarray}
\frac{\partial V}{\partial \Phi}|_{t=t_z}=0
\label{res9012}
\end{eqnarray}
In Eq. (\ref{res9012})  all couplings are known except for $\lambda$. We further require that the minimum is obtained for $\Phi=v=246.22$ GeV and solve the minimum equation for the parameter $\lambda$. Here we make the underlying assumption that if the potential is phenomenologically viable as an effective potential then it should lead to  a mass of the Higgs boson very close to the actual mass (This is actually exact in the on-shell subtraction scheme where the renormalized mass is equal to the pole one).
\begin{figure}
\begin{center}
\epsfxsize = 10cm
\epsfbox{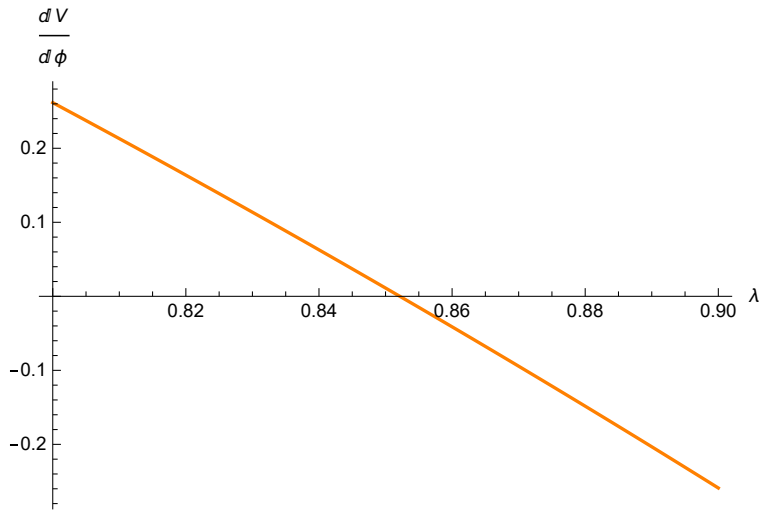}
\end{center}
\caption[]{%
Plot of $\frac{\partial V}{\partial \Phi}|_{\Phi=v}$ as function of the parameter $\lambda$ in the effective potential V.
}
\label{min}
\end{figure}

In Fig. \ref{min} we plot $\frac{\partial V}{\partial \Phi}|_{\Phi=v}$ in terms of the parameter $\lambda$ to determine $\lambda=0.852$. We use this value further calculate:
\begin{eqnarray}
\frac{\partial^2 V}{\partial \Phi^2}=m_h^2.
\label{mass7564}
\end{eqnarray}
Then the resulting mass of $m_h=126.15$ GeV is very close to the actual experimental mass of the Higgs boson $m_{hexp}=125.09$ GeV. By varying the top Yukawa coupling (here we took $y=\frac{m_t}{\sqrt{2}v}$ with the mass of the top quark $m_t=174.135$ GeV \cite{PDG}) one can reproduce the exact pole mass of the Higgs boson.

\section{Higgs effective couplings to two photons and two gluons}
An unusual feature of the potential obtained in section II and particularized in section III is that it contains terms of the type $\ln[\frac{G^{\mu\nu}G_{\mu\nu}}{\Lambda^4}]$ (exemplified here for $U(1)_Y$). These terms are finally integrated out. The logarithms introduce singularities if the fields are close to zero but we fixed the scale of our potential the electroweak scale so we expect values of the fields around that scale. For all range of values however one would need to regularize the corresponding terms. Things can also be regarded differently. Depending on the nature of their beta functions the couplings of the gauge fields may become strong at higher or lower scales. Then it is possible that at that scale a phase transition occurs and gauge condensates forms. Neglecting the anomalous dimensions of the gauge fields then it make sense to expand the logarithms against the scale where the coupling become strong which virtually would identify with the scale of the condensate. This approach which we will consider here is very helpful in determining the Higgs couplings to two photons and two gluons for the potential in section III.

To illustrate this we first consider the decay of the Higgs boson to two photons discussed in detail in the literature \cite{Ellis}, \cite{Vainshtein}, \cite{Marciano}. For a Higgs coupling with two photons of the type:
\begin{eqnarray}
\frac{1}{2}F_hF^{\mu\nu}F_{\mu\nu}h
\label{higgr}
\end{eqnarray}
where $F^{\mu\nu}$ is the electromagnetic tensor, $h$ is the Higgs boson and $F_h$ is the coupling, the amplitude of the two photon decay of the Higgs is \cite{Vainshtein}:
\begin{eqnarray}
A(h\rightarrow\gamma\gamma)=F_h(k_{1\mu}\epsilon_{1\nu}-k_{1\nu}\epsilon_{1\mu})(k_{2\mu}\epsilon_{2\nu}-k_{2\nu}\epsilon_{2\mu}),
\label{amp884664}
\end{eqnarray}
where $k_1$, $k_2$ are the momenta of the two photons and $\epsilon_1$, $\epsilon_2$ are their polarizations.
In the standard model at one loop \cite{Marciano}:
\begin{eqnarray}
F_h=\frac{e^2g}{(4\pi)^2m_W}\frac{1}{2}F,
\label{coupl64788364}
\end{eqnarray}
where,
\begin{eqnarray}
F=F_W(\beta_W)+\sum_fN_cQ_f^2F_f(\beta_f).
\label{funct5674}
\end{eqnarray}
Here $N_c$ is the color factor ($N_c=2$ for leptons and $N_c=3$ for quarks) and,
\begin{eqnarray}
&&\beta_W=\frac{4m_W^2}{m_h^2}
\nonumber\\
&&\beta_f=\frac{4m_f^2}{m_h^2}.
\label{res8564788}
\end{eqnarray}
Furthermore,
\begin{eqnarray}
&&F_W(\beta)=2+3\beta+3\beta(2-\beta)f(\beta)
\nonumber\\
&&F_f(\beta)=-2\beta[1+(1-\beta)f(\beta)]
\label{detfunc65774},
\end{eqnarray}
with,
\begin{eqnarray}
&&f(\beta)=\arcsin^2(\beta^{-\frac{1}{2}})\,\,\,{\rm for}\beta\geq 1
\nonumber\\
&&f(\beta)=-\frac{1}{4}\Bigg[\ln[\frac{1+\sqrt{1-\beta}}{1-\sqrt{1-\beta}}]-i\pi\Bigg]^2
\label{thridfunc64537}
\end{eqnarray}
For the values of the parameters at the electroweak scale and considering only the top quark the couplings $F$ (of Higgs to two photons) and $F_t$ (of Higgs to two gluons) have the values;
\begin{eqnarray}
&&F_t=-1.376
\nonumber\\
&&F=6.5
\label{coupl5674884}
\end{eqnarray}

For comparison we will determine the Higgs couplings in our potential before integrating out the gauge fields. We will explain in detail how this works for the decay to two photons of the Higgs boson and apply briefly
our results to the two gluon decay of the Higgs boson because the results are very similar.
The relevant term in the Lagrangian is:
\begin{eqnarray}
&&{\cal L}_s=-b_1G^{\mu\nu}G_{\mu\nu}\ln[\frac{G^{\mu\nu}G_{\mu\nu}}{4m_Z^4}]-
-b_2G^{\mu\nu}G_{\mu\nu}\ln[\frac{\Phi^4}{4m_Z^4}]-
\nonumber\\
&&c_1F^{a\mu\nu}F^a_{\mu\nu}\ln[\frac{F^a_{\mu\nu}F^{a\mu\nu}}{4m_Z^4}]-
c_2F^{a\mu\nu}F^a_{\mu\nu}\ln[\frac{\Phi^4}{4m_Z^4}].
\label{relpot677}
\end{eqnarray}
We expand around the Higgs vev:
\begin{eqnarray}
\ln[\frac{\Phi^4}{4m_Z^4}]=\ln[\frac{(v+h)^4}{4m_Z^4}]=\ln[\frac{v^4}{4m_Z^4}]+\frac{h}{v}+...
\label{res765789}
\end{eqnarray}
where we detained only the relevant terms.
We use:
\begin{eqnarray}
&&G^{\mu\nu}G_{\mu\nu}=\cos^2\theta_WF^{\mu\nu}F_{\mu\nu}+...
\nonumber\\
&&F^{a\mu\nu}F^a_{\mu\nu}=\sin^2\theta_WF^{\mu\nu}F_{\mu\nu}+...,
\label{expi76859}
\end{eqnarray}
where$F^{\mu\nu}$ is the electromagnetic tensor.

The logarithms of the gauge fields are then expanded around the scale where the coupling constant is strong which is $g^{\prime2}=1$ for the $U(1)_Y$ group (such that $e^2_a=cos^2\theta_Wg^{\prime2}=\cos^2\theta_W$, where $e^2_a$ is the electric charge at that scale) and $g^2=1$ for $SU(2)_L$ (such that $e^2_b=\sin^2\theta_W$,  where $e^2_b$ is the electric charge at the second scale).
One can use the beta function for the electromagnetic coupling,
\begin{eqnarray}
\frac {d e}{d \ln(\mu)}=\frac{1}{16\pi^2}\frac{11}{3}e^3,
\label{betf76885}
\end{eqnarray}
to integrate it out,
\begin{eqnarray}
\frac{1}{2e_Z^2}-\frac{1}{2e_1^2}=\frac{1}{16\pi^2}\frac{11}{3}\ln[\frac{\mu_1}{\mu_Z}],
\label{int74665}
\end{eqnarray}
where $\mu_Z=m_Z$ is the electroweak scale and $\mu_1$ is the strong coupling scale to find:
\begin{eqnarray}
&&\ln[\frac{\mu_{1a}^4}{m_Z^4}]=\frac{6}{11}16\pi^2[\frac{1}{e_z^2}-\frac{1}{\cos^2\theta_W}]
\nonumber\\
&&\ln[\frac{\mu_{1b}^4}{m_Z^4}]=\frac{6}{11}16\pi^2[\frac{1}{e_z^2}-\frac{1}{\sin^2\theta_W}],
\label{res6645}
\end{eqnarray}
where $\mu_{1a}$ is the scale where $g^{\prime 2}\approx 1$ and $\mu_{1b}$ is the scale where $g^2\approx 1$. Then one can rewrite Eq. (\ref{relpot677}) as:
\begin{eqnarray}
&&{\cal L}_S=-b_1\cos^2\theta_WF^{\mu\nu}F_{\mu\nu}\ln[\frac{\mu_{1a}^4}{m_z^4}]-c_1\sin^2\theta_WF^{\mu\nu}F_{\mu\nu}\ln[\frac{\mu_{1b}^4}{m_z^4}]+
\nonumber\\
&&-b_2F^{\mu\nu}F_{\mu\nu}\ln[\frac{v^4}{4m_Z^4}]+b_2F^{\mu\nu}F_{\mu\nu}\frac{h}{v}+....
\label{finalexpr746353}
\end{eqnarray}
where we omitted the irrelevant terms which are assimilated with interactions. Here we also used the fact that $b_2=c_2$.

Consequently the kinetic terms for the electromagnetic field receives a factor $s$,
\begin{eqnarray}
&&s=4b_1\cos^2\theta_W\frac{6}{11}16\pi^2[\frac{1}{e_Z^2}-\frac{1}{\cos^2\theta_W}]+
\nonumber\\
&&4c_1\sin^2\theta_W\frac{6}{11}16\pi^2[\frac{1}{e_Z^2}-\frac{1}{\sin^2\theta_W}]+4b_2\ln[\frac{v^4}{4m_Z^4}],
\label{res62884993}
\end{eqnarray}
which leads to an amplitude of  Higgs decay to two photons is:
\begin{eqnarray}
A_V(h\rightarrow\gamma\gamma)=\frac{-8b_2}{vs}(k_{1\mu}\epsilon_{1\nu}-k_{1\nu}\epsilon_{1\mu})(k_{2\mu}\epsilon_{2\nu}-k_{2\nu}\epsilon_{2\mu}).
\label{ampnew64774}
\end{eqnarray}
From Eqs. (\ref{amp884664}), (\ref{coupl64788364}) and (\ref{ampnew64774}) we obtain the correspondence:
\begin{eqnarray}
\frac{-8b_2}{s}\rightarrow F
\label{res7395784}
\end{eqnarray}
where for our model,
\begin{eqnarray}
\frac{-8b_2}{s}=6.821.
\label{finalres78888}
\end{eqnarray}
The result in Eq. (\ref{finalres78888}) is very close to the value for the one loop in the standard model computed in Eq. (\ref{coupl5674884}).

The same method can be applied to the two gluon decay of the Higgs boson with the provision that as the scale of reference $\Lambda_{QCD}$ such that,
\begin{eqnarray}
\ln\frac{\Lambda_{QCD}^4}{m_Z^4}=16\pi^2(-\frac{7}{2}\frac{1}{g_{3Z}^2}),
\label{resgl75886}
\end{eqnarray}
where $g_{3Z}$ is the strong coupling constant at the electroweak scale $m_Z$. The kinetic term for the gluon field will receive in front a factor:
\begin{eqnarray}
s'=4d_1(-\frac{2}{7}16\pi^2\frac{1}{g_{3Z}^2})+4d_2\ln[\frac{v^4}{4m_Z^4}].
\label{res784ygl}
\end{eqnarray}
The amplitude of Higgs decaying to two gluons is,
\begin{eqnarray}
A(h\rightarrow gg)=\frac{-8d_2}{s'v}(k_{1\mu}\epsilon^a_{1\nu}-k_{1\nu}\epsilon_{1\mu})(k_{2\mu}\epsilon^a_{2\nu}-k_{2\nu}\epsilon_{2\mu}).
\label{res664}
\end{eqnarray}
Here  $k_1$ and $k_2$ are the momenta of the two gluons and  $\epsilon^a_1$ and $\epsilon^a_2$ are their polarizations. Then from Eqs.  (\ref{coupl64788364}) and (\ref{res664}) the following correspondence is obtained:
\begin{eqnarray}
\frac{-8d_2}{s'}\rightarrow f_2,
\label{cnut710}
\end{eqnarray}
where for our model $\frac{-8d_2}{s'}=-1.091$  again very close to the standard model value computed at one loop in Eq. (\ref{coupl64788364}).

In this section we computed the decay widths to two photons and two gluons of the Higgs boson in the context of an effective model. These decay widths as stated here depend on four parameters $b_2$, $d_2$, $s$ and $s'$ which in their turn depend on the top Yukawa and the gauge coupling constants. Again from low energy QCD \cite{Jora6}, \cite{Jora7} we learn that in the context of an effective theory one should not expect that the parameters that describe the effective widths should be directly related to those employed in standard tree level or one loop calculations or to those employed in other types of models that describe the same processes. For a phenomenological model to be viable it is necessary and sufficient only that the phenomenological result to be aligned to the experimental data or other consistent theoretical results.

\section{Conclusions}

In this work we proposed an effective Higgs model constructed not by integrating out at one or two loops the gauge, fermion and scalar degrees of freedom but by analogy with the low energy QCD linear sigma models. Thus this kind of model may be suitable both for the case when the Higgs boson is elementary and also when it is the result of some unknown strong dynamics. All terms in this potential apart from the $\lambda$ term are directly derived from the trace anomaly expressed as the product between the beta function of the coupling constant and the two dimensional or four dimensional operators characterized by it. First we constructed a general potential that contained a number of $20$ parameters along with 5 constraints. In this case available phenomenological data might be used to fix the parameters.

 Next we consider a particular case of the potential again inspired by the construction of low energy QCD effective models \cite{Schechter10} and also from the construction of the standard one loop effective Higgs models \cite{Casas1}, \cite{Casas2}. Moreover we eliminated all redundant parameters. We fixed the scale of our model at $\mu=m_Z$ and applied the minimum condition
$\frac{\partial V}{\partial\Phi}=0$  for $\Phi=v$ to determine the value of the quadrilinear coupling constant $\lambda$. With this value we further on computed the Higgs mass as $\frac{\partial^2 V}{\partial \Phi^2}=m_h^2$.
Our result of $m_h=126.15$ GeV agreed well with the known experimental value of the Higgs boson $m_{hexp}=125.09\pm0.24 $ GeV.

In the same framework but before integrating out the gauge degrees of freedom we determined the effective Higgs couplings to two photons and two gluons again in good accordance with what we know form one loop calculations. { This shows  that the particular case of the model we proposed already describes very well at least a few phenomenological quantities.  The potential derived here can be used to extract other possible couplings along the same lines.

 The presence of  a large number of parameters in the most general version of the model constructed here should not be regarded as a lack of predictability as compared to standard calculations at some loop order of the effective potential but as a way of encapsulating our lack of knowledge with regard to higher loop corrections to the phenomenological parameters. Since in general it is easier to compute beta functions and anomalous dimensions than intricate processes this kind of model, especially if one uses physical arguments to further constrain or determine some of the parameters, may have important applications.

One potential application of our model would be to study the vacuum stability of the standard model. This topic was thoroughly studied in the framework of regular one loop or two loop renormalization improved effective potentials \cite{Jones1}, \cite{Jones2}. Some authors \cite{Schwartz1}, \cite{Schwartz2} argued that since the actual mass of the Higgs boson discovered at the LHC situates the standard model at the border between absolute stability and metastability the lack of gauge invariance of usual effective model may play negative role in this issue. Of course any effective potential expressed in terms of the classical field is non gauge invariant but any physical quantities obtained from should be. The potential constructed here not only satisfies in detail the trace anomaly but also in its primitive form before integrating out the gauge, fermion, scalars gauge degrees of freedom has all the terms gauge invariant again apart from the classical field. Then one can extract useful gauge invariant effective couplings of the Higgs boson with the other fields in the standard model Lagrangian.

Other possible aspects and application of our method will be investigated in further work.

\end{document}